\documentclass[fleqn,10pt]{wlscirep}
\usepackage[utf8]{inputenc}
\usepackage[T1]{fontenc}

\usepackage{ragged2e}

\usepackage{threeparttable}
\usepackage{multirow}
\usepackage{longtable}
\usepackage{rotating}
\usepackage{soul}
\usepackage{graphicx}
\usepackage{subfigure} 
\usepackage{tabularx,colortbl}
\usepackage{float}
\usepackage{cite}

\usepackage{amsmath,amssymb,amsthm}

\usepackage{xcolor}

\title{Extremely asymmetric absorption and reflection near the exceptional point of three-dimensional metamaterial}

\author[1,2]{Yanjie Wu}
\author[3]{Ding Zhang}
\author[1]{Qiuyu Li}
\author[1,*]{Hai Lin}
\author[1]{Xintong Shi}
\author[1,4]{Jie Xiong}
\author[2]{Haoquan Hu}
\author[2]{Jing Tian}
\author[3]{Bian Wu}
\author[5,*]{Y. Liu}

\affil[1]{College of Physics Science and Technology, Central China Normal University, Wuhan 430079, Hubei Province, People’s Republic of China}

\affil[2]{School of Electronic Science and Engineering, University of Electronic Science and Technology of China, Chengdu 611731, Sichuan Province, People’s Republic of China}

\affil[3]{National Key Laboratory of Antennas and Microwave Technology, Xidian University, Xi’an 710071, Shaanxi Province, People’s Republic of China}

\affil[4]{School of Physics and Telecommunications, HuangGang Normal University, Huanggang 438000, Hubei Province, People’s Republic of China}

\affil[5]{Department of Physics, School of Physics, Hubei University, Wuhan 430062, Hubei Province, People’s Republic of China}

\affil[*]{corresponding. linhai@mail.ccnu.edu.cn, yangjie@hubu.edu.cn}

\begin{abstract}
In recent years, particular physical phenomena enabled by non-Hermitian metamaterial systems have attracted significant research interests. In this paper, a non-Hermitian three-dimensional metamaterial near the exceptional point (EP) is proposed to demonstrate extremely asymmetric absorption and reflection. Unlike its conventional counterparts, this proposed metamaterial is constructed with a loss-assisted design. Localized losses are introduced into the structure by combining our technique of graphene-based resistive inks with conventional printed circuit board (PCB) process. Extremely asymmetric absorption and reflection near the EP are experimentally observed by tuning the loss between split ring resonators (SRRs) in the meta-atoms. Simultaneously, by linking the equivalent circuit model (ECM) with the Hamiltonian quantum physical model, the equivalent non-Hermitian Hamiltonian is obtained and a non-Hermitian transmission matrix is constructed. We show that tuning the structure and circuit parameters of the ECM produces a metamaterial system with EP response. Our system can be used in the design of asymmetric metamaterial absorbers. Our work lays down the way for the manipulation of EP to develop perfect absorption, sensing and other applications in the 3D metamaterial platform.
\end{abstract}
\begin{document}

\flushbottom
\maketitle
%
%


\section*{Introduction}

Metamaterial is an artificial composite material made of sub-wavelength scale man-made structures arranged in periodic or non-periodic arrangements, which has unique properties not found in natural materials~\cite{RN102,RN273}. Previous metamaterials have been considered as an energy conservation system~\cite{RN101}. These wave systems can be naturally described by Hermitian Hamiltonians since the energy dissipation has been omitted~\cite{RN103}. However, in many applications, the loss (e.g., radiation loss, ohmic loss) is unavoidable in metamaterial designs due to the interactions between lossy/active meta-atoms and the environment~\cite{RN955}. In a broad sense, all open systems that exchange energy with the environment are non-Hermitian systems. In recent years, with the development of non-Hermitian physics, these open systems are associated with non-Hermitian Hamiltonians~\cite{RN104}. The non-Hermiticity thus results in extending the effective permittivity and permeability of metamaterials to the complex plane. In these non-conservative situations, the rich physics of non-Hermitian systems have been revealed. As a result, a series of attractive applications of counter-intuitive wave phenomena have been unlocked, such as laser-absorber \cite{RN105}, unidirectional reflectionless transmission \cite{RN107}, extreme asymmetric reflection/absorption \cite{RN109}, single laser-mode selection \cite{RN111}. The exceptional point (EP) is one ubiquitous concept in non-Hermitian physics which was first proposed by Carl Bender and Stefan Boettcher \cite{RN113}. It has attracted a lot of research interest \cite{RN114,RN115}. The EPs are non-Hermitian energy degeneracies at which the eigenvalues and corresponding eigenvectors simultaneously coalesce \cite{RN117}. Near the EP of the parameter space, the eigenvalue spectrum of these systems from real-valued becomes complex due to its non-Hermitian nature \cite{RN118}. The wave behaviors near EPs have resulted in demonstrations of interesting phase transitions, such as unidirectional invisibility \cite{RN119}, ultra-sensitive \cite{RN121}, loss-induced transparency \cite{RN123}, optical isolation \cite{RN125} and non-reciprocity \cite{RN127}. These interesting features could be key to expanding the application scopes of metamaterials.

Due to the existence of EP-induced phenomena, EPs have received much attention in the field of photonics and acoustics in non-Hermitian Parity-Time ($PT$) symmetric systems \cite{RN110,RN930_45}. $PT$ Symmetry means the invariance of the Hamiltonian under the operations of space inversion and time inversion \cite{RN948}. To realize $PT$ symmetry in non-Hermitian electromagnetic (EM) systems, a wide employment scheme is to intentionally balance gain and loss \cite{RN103}. However, in practical applications, it can be cumbersome to introduce and manipulate gain factors in metamaterial platforms and to balance them with loss factors \cite{RN957_40,RN938_41}. Recent research shows that all-passive systems without optical gain can be engineered to demonstrate EP-related phenomena \cite{RN108,RN129,RN130}. Li et al.~\cite{RN110} demonstrated a passive asymmetric acoustic absorber at an EP with a compact configuration and deep-subwavelength thickness which achieves EP-induced tunable asymmetric absorption. Gu et al.~\cite{RN135} proposed a non-ideal $PT$ metasurface, which realizes controllable unidirectional reflectionless propagation at EP by simply adjusting the angle of the incident wave as well as the structure parameter of the meta-atom. Correspondingly, the EP can occur in the resonant asymmetric metasurface which is designed based on bianisotropy \cite{RN919_42}. Moreover, Dong et al.~\cite{RN136} proposed to utilize the loss to construct non-Hermitian EM metasurfaces. The designed metasurface enables extraordinary angular asymmetry at EP condition. Similarly, the introduction of absorptive defects in local regions can convert an ordinary symmetric retroreflector into an extraordinary asymmetric one \cite{RN953_44}. However, the non-Hermitian system in these previous works only provides a platform to validate general concepts in EP physics. It appears to have limited application situations or depend more on the laboratory environment \cite{RN922_46,RN923_47}. The EP theory has not been used to guide the design of practical functional metamaterial EM devices.

Since the equivalent circuit model (ECM) can accurately combine the geometric parameters with the circuit parameters, the equivalent circuit theory is a common method to analyze perfect metamaterial absorbers \cite{RN155_48,RN836_49}. In recent years, several scholars have explored the non-Hermitian phenomena of the system using ECMs \cite{RN106, RN966_50}. In \cite{RN936}, a platform for exploring the topological features of bound states in the continuum (BIC) and related phenomena was formed using radio-frequency (RF) electronic circuit. It forms an EP of zeros and realizes unusual physics by merging two coherent perfect absorption (CPA) solutions. However, it is only verified from the perspective of the circuit, and no actual structural model is designed through the RF circuit to establish the correspondence between the dimensional parameters of the structure and the circuit parameters. Moreover, in our previous studies \cite{RN43}, the absorption mechanism of the three-dimensional metamaterial absorber (3D-MA) is only analyzed by normalized impedance and surface current distribution. Therefore, it is valuable to explore the properties of metamaterial absorbers (MAs) with extremely asymmetric absorption from the perspective of non-Hermitian physics \cite{RN925}.

In this work, we propose a non-Hermitian three-dimensional metamaterial (3D-MM) by manipulating the local loss distribution. The proposed metamaterial can be used as a unidirectional absorber which processes an absorption band around the EPs. Previously, the conventional design theories of MAs were based on impedance matching theory \cite{RN298,RN390}. The MAs are considered as a single port network since there are metal plates on their backside which block any backward incident wave. While in this design, the metamaterial is assembled into hollow mesh array form, and the metal-back plate is removed to allow energy leakage to the opposite side of the incoming wave. At the same time, the unique design of the mesh array structure without the metal-back plate enables the metamaterial to exhibit optically transparent properties. The graphene-based resistive film is used to introduce the localized loss in structures. By tuning the loss between split ring resonators (SRRs) in the meta-atoms, we observe extremely asymmetric absorption and reflection relevant to EPs. The metamaterial system achieves a quasi-perfect absorption ($absorptivity \ge 99\%$) for forward incidence and near-total reflection for opposite incidence in the designated frequency band. Moreover, this system constructs a low insertion loss transmissive window out of the operating frequency. Different from previous works, this paper denotes the S-parameter of the modeled two-port network of the whole system as the Hamiltonian matrix of non-Hermitian form. The eigenvalue surface of the non-Hermitian matrix forms a self-intersecting Riemann surface whose EP is located at the intersection. We combine the equivalent circuit model with the Hamiltonian quantum physical model to construct the non-Hermitian transmission matrix of the designed structure, thus establishing a new method for designing metamaterial systems with EP response using the equivalent circuit design approach. In this manner, we apply the EP theory in non-Hermitian physics to the design of practical functional metamaterial devices.

\section*{Theory}

\begin{figure*}[!ht]
  \centering\includegraphics[width=6in]{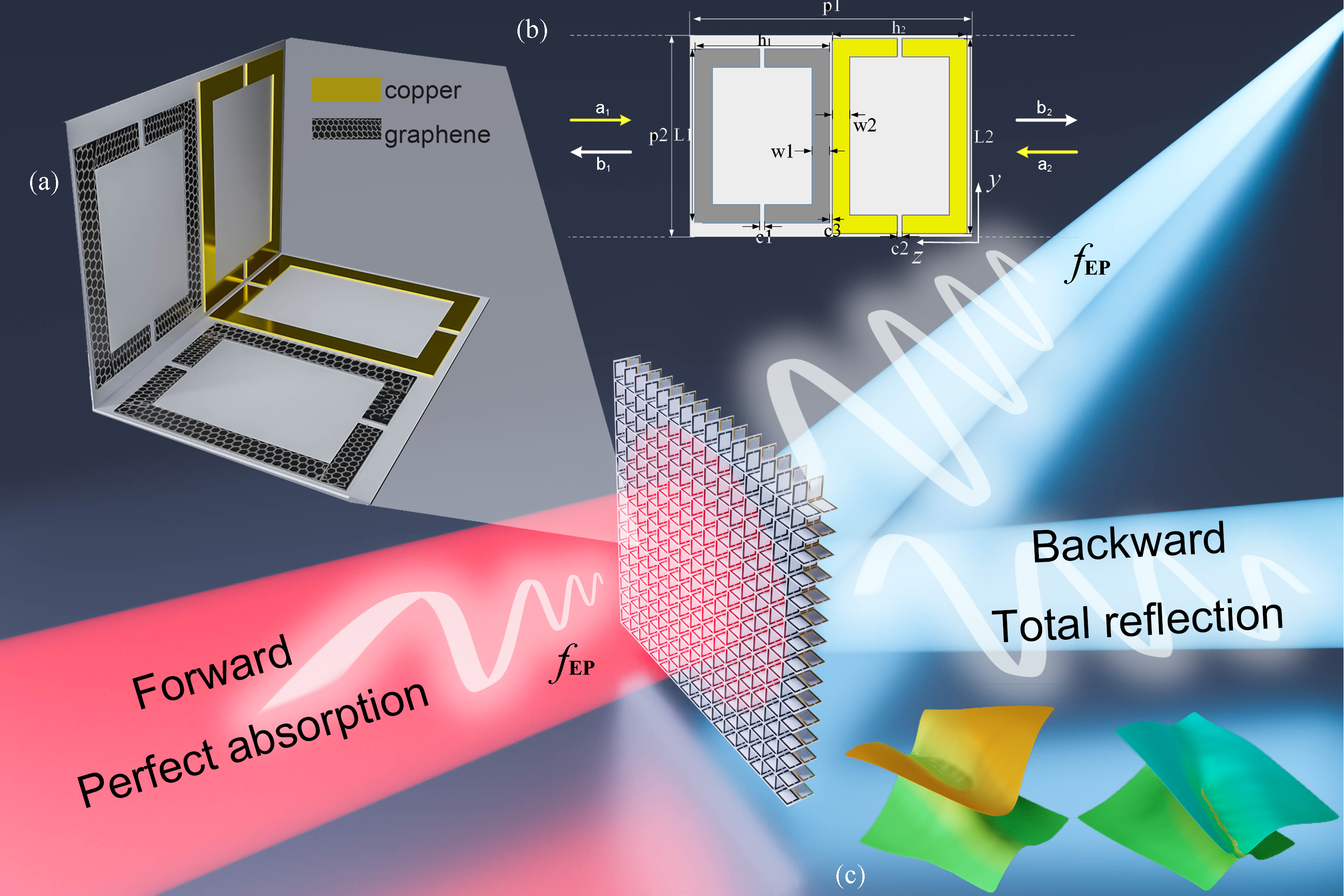}
 \caption{Conceptual illustration of asymmetric absorption and reflection near the EP of the designed 3D-MM. (a) 3D illustration of unit cell. (b) Two-dimensional view of the designed 3D-MM. (c) Schematic diagram of the real part and imaginary part of eigenvalue surfaces forming self-intersecting Riemann surfaces when EP exists.}
  \label{Fig1}
\end{figure*}

The concept diagram of the extremely asymmetric absorption and reflection of the designed un-reciprocal 3D-MM near the EP is illustrated in Figure 1. Therein, the colors of the beams stand for incidences in different directions for our designed 3D-MM. The red beam exhibits perfect absorption near the EP for the forward incidence, and the blue beam for the backward incidence when the total reflection is achieved near the EP. The designed 3D-MM is composed of a hollow vertical periodic crossed mesh array. As shown in inset (a) of Figure 1, the lossy graphene-based resistive film and metallic copper are introduced to realise the extremely asymmetric absorption and reflection phenomenon near the EP. Inset (b) of Figure 1 shows the design parameter definition of the meta-atom in the proposed design. From Figure 1, it can be seen that the 3D-MM consists of two SRRs with different sizes, which are made of different materials as follows. The larger SRR is made of metal copper, and the smaller SRR is made of the graphene-based resistive film with a sheet resistance of 30 $\Omega/sq$. Since the loss of the impedance film is uniform, we use graphene-based resistive film to replace the lumped element as the loss part. The SRRs were printed on 0.813 mm thick Rogers RO4003C. The relative permittivity of the substrate is 3.38 and the loss tangent is 0.0027. Inset (c) of Figure 1 shows the self-intersecting Riemann surface formed by the real and imaginary parts of the eigenvalues when EP exists. 

Non-Hermitian metamaterial systems can be investigated by scattering matrix theory to examine the interaction of metamaterials with external EM waves. In order to verify the extremely asymmetric absorption and reflection phenomenon of the designed 3D-MM, the metamaterial system is modelled as a two-port network. Analogously to the Hamiltonian matrix $H$ in quantum mechanics, the eigenvalues of designed 3D-MM can be solved from the scattering matrix. By exploring the subspace of eigenvalues and the Riemann surface distribution, the possibility of constructing non-Hermitian EP systems based on the 3D-MM platform can be explored theoretically.

As shown in inset (b) of Figure 1, suppose that the designed 3D-MM is a two-port network \cite{RN826}, the wave propagation in the system can be described by the scattering matrix $\mathbf{S}_m$. The forward (port 1) and backward (port 2) incidences of EM waves are represented by $a_1$ and $a_2$, respectively. The corresponding reflected waves are represented by $b_1$ and $b_2$. Their quantitative relationship can be expressed as

\begin{equation}
\left[\begin{array}{l}b_1 \\ b_2\end{array}\right]=\mathbf{S}_m\left[\begin{array}{l}a_2 \\ a_1\end{array}\right],   \mathbf{S}_m=\left[\begin{array}{cc}S_{12} & S_{11} \\ S_{22} & S_{21}\end{array}\right]=\left[\begin{array}{cc}T_b & R_f \\ R_b & T_f\end{array}\right], 
\end{equation}
where $S_{12}$ ($T_b$), $S_{21}$ ($T_f$) represents the transmission coefficient, and  $S_{11}$ ($R_f$),  $S_{22}$ ($R_b$) represents the reflection coefficient at forward and backward incidence, respectively. However, since the designed metamaterial system is a multilayer structure, it can be represented as a cascade of multiple two-port networks. We relate the field on one side of the 3D-MM to the field on the other side by a transfer matrix. The transfer matrix can be used to calculate the various parameters of the scattering matrix \cite{RN302,RN303,RN851}. The total transfer matrix of the designed metamaterial system can be defined as

\begin{equation}
\left[\begin{array}{l}E_f \\H_f
\end{array}
\right]=\mathrm{\mathbf{T}}_{\text {total}}\left[\begin{array}{l}
E_b \\
H_b
\end{array}\right], \mathrm{\mathbf{T}}_{\text {total }}=\left[\begin{array}{cc}
A & B \\
C & D
\end{array}\right]
\end{equation}
where $E_f$ and $H_f$ are the electric field and magnetic field in the forward direction of 3D-MM, and $E_b$ and $H_b$ are the electric field and magnetic field in the backward direction of the 3D-MM. In addition, the total transfer matrix of the cascaded network can be represented by the ABCD matrix.

According to the transformation relation between the transfer matrix and scattering matrix and the basic definitions of scattering matrix \cite{RN827,pozar2011microwave}, the reflection and transmission coefficients can be expressed as

\begin{equation}
S_{11}=\frac{A Z_1+B-C Z_1^2-D Z_1}{A Z_1+B+C Z_1^2+D Z_1}
\end{equation}

\begin{equation}
S_{21}=\frac{2Z_1}{A Z_1+B+C Z_1^2+D Z_1}
\end{equation}
where $Z_1$ denotes the characteristic impedance of the transmission line. By the calculation of Equation (3) and Equation (4), we can obtain the reflection coefficient $S_{11}$ ($R_f$) and transmission coefficient $S_{21}$ ($T_f$) when the EM wave is incident from forward. When the EM wave is incident from backward, according to Equation (2), the following equation can be obtained

\begin{equation}
\left[\begin{array}{l}
E_b \\
H_b
\end{array}\right]=\left[\begin{array}{cc}
A & B \\
C & D
\end{array}\right]^{-1}\left[\begin{array}{l}
E_f \\
H_f
\end{array}\right]
\end{equation}

Similarly, Equation (3) and Equation (4) can be used for similar derivation, so as to obtain the reflection coefficient $S_{22}$ ($R_b$) and the transmission coefficient $S_{12}$ ($T$) when the EM wave is incident from port 2.

\begin{equation}
S_{22}=\frac{-A Z_1-B+C Z_1^2+D Z_1}{A Z_1+B+C Z_1^2+D Z_1}
\end{equation}

\begin{equation}
S_{12}=\frac{2det(T_{total})Z_1}{A Z_1+B+C Z_1^2+D Z_1}
\end{equation}

\begin{figure*}[!ht]
  \centering\includegraphics[width=6in]{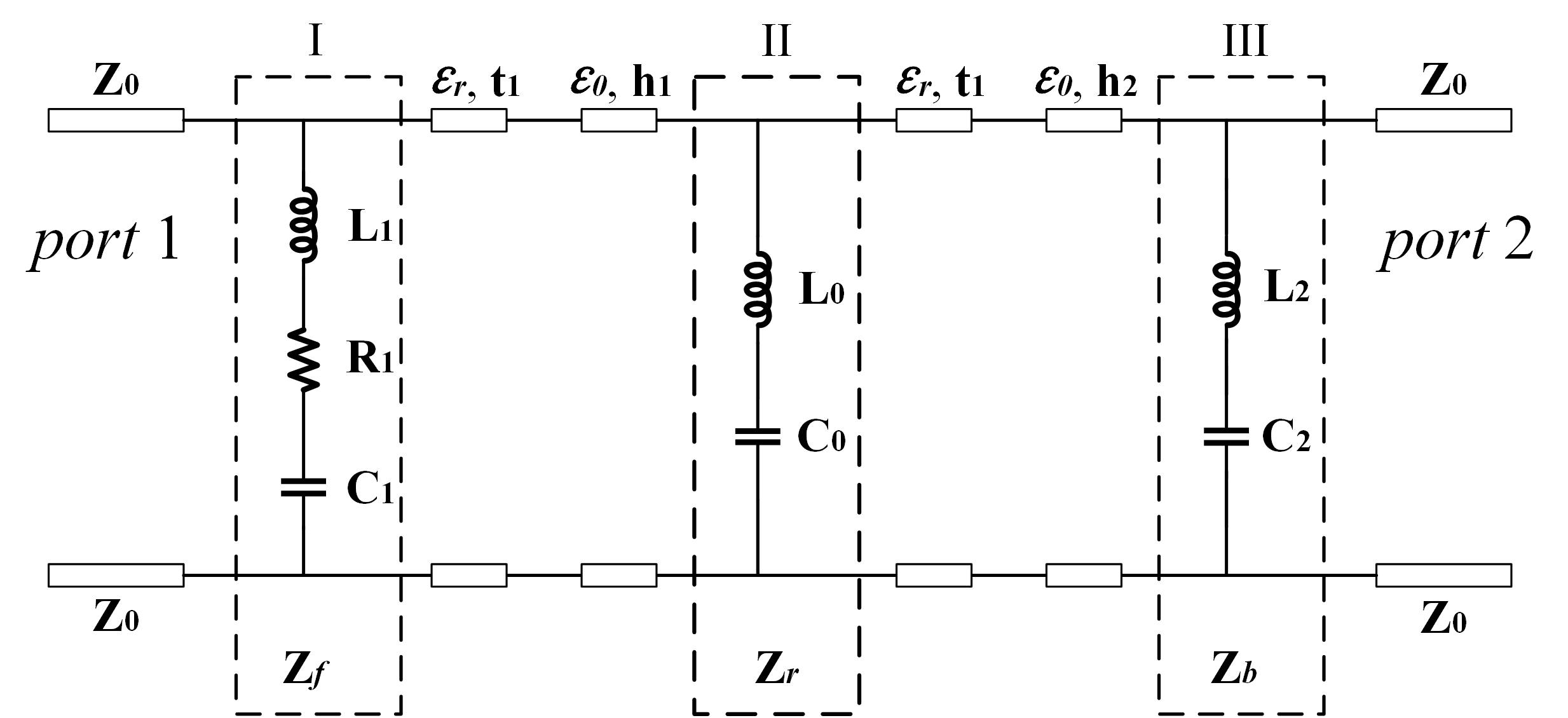}
 \caption{Equivalent circuit model of the designed 3D-MM.}
  \label{Fig2}
\end{figure*}

At this point, we obtained the scattering matrix $\mathbf{S}_m$ of the designed 3D-MM. Moreove, since ECM is able to represent abstract physical concepts with detailed circuit parameters in a more visual and clear way from an engineering perspective, we use equivalent circuit theory to analyze the designed metamaterial system. The ECM of the designed structure can also be denoted by the transfer matrix. Therefore, it is possible to relate the $S$-parameters of the system with the ECM.

Due to the fact that the designed 3D-MM is of the hollow mesh array structure, it is difficult to extract its exact one-to-one corresponding ECM \cite{RN44}. Therefore, we regard it as a two-dimensional structure and extract its approximate ECM as shown in Figure 2. The designed 3D-MM can be divided into three parts, $Z_f$ is the equivalent impedance of the smaller SRR. $Z_b$ is the equivalent impedance of the larger SRR. $Z_r$ is the equivalent impedance between the larger SRR and the smaller SRR. $Z_0$ is the wave impedance in free space. $L$ and $C$ are the equivalent resistance and equivalent capacitance of each part of the structure. $\varepsilon_0$ and $\varepsilon_r$ are the relative permittivity of free space and dielectric substrate, respectively. $t_1$ is the thickness of the substrate. $h_1$ and $h_2$ are the widths of the larger SRR and smaller SRR.

According to the characteristics of the two-port network and the ECM of the designed 3D-MM, the total transfer matrix of the designed 3D-MM can be obtained by multiplying the transfer matrix for each component, expressed as

\begin{equation}
\mathrm{\mathbf{T}}_{\text {total }}=\left[\begin{array}{ll}
A & B \\
C & D
\end{array}\right]=\mathbf{M}_f \mathbf{M}_1 \mathbf{M}_0 \mathbf{M}_2 \mathbf{M}_b, 
\end{equation}
where $\mathbf{M}_f$, $\mathbf{M}_1$, $\mathbf{M}_0$, $\mathbf{M}_2$ and $\mathbf{M}_b$, represent the transfer matrix at region I, between region I and region II, at region II, between region II and region III, and at region III, respectively. 

The transfer matrix of each component will be derived individually. Similar to Equation (2), the fields on both sides of each component can be related. The transfer matrix of the region I in Figure 2 can be expressed by the following equation
\begin{equation}
\left[\begin{array}{l}
E_1 \\
H_1
\end{array}\right]=\mathbf{M}_f\left[\begin{array}{l}
E_2 \\
H_2
\end{array}\right],  \mathbf{M}_f=\left[\begin{array}{cc}
1 & 0 \\
\frac{1}{z_f} & 1
\end{array}\right]
\end{equation}
where $\mathrm{Z}_f=j \omega L_1+\frac{1}{j \omega C_1}+R_1$,  $\left[\begin{array}{l}
E_1 \\
H_1
\end{array}\right]$ and $\left[\begin{array}{l}
E_2 \\
H_2
\end{array}\right]$ are the electric field and magnetic field on the left and right sides of the region I, respectively. Equation (9) regards the region I as a two-port network, and the system resonance and loss distribution of the region I can be effectively manipulated by adjusting the values of $L_1$, $C_1$ and $R_1$. Likewise, the transfer matrix between region I and region II are

\begin{equation}
\left[\begin{array}{l}
E_2 \\
H_2
\end{array}\right]=\mathbf{M}_1\left[\begin{array}{l}
E_3 \\
H_3
\end{array}\right], \mathbf{M}_1=\left[\begin{array}{cc}
\cos \beta h_1 & j Z_0 \sin \beta h_1 \\
j \frac{1}{z_0} \sin \beta h_1 & \cos \beta h_1
\end{array}\right]
\end{equation}
where $\beta=2 \pi f / c=\omega \sqrt{\mu_0 \varepsilon_0}$, $\beta$ is the propagation constant of free space, $\omega$ is the angular frequency, $\mu_0$ and $\varepsilon_0$ are the permeability and permittivity in free space respectively, $Z_0$ is the impedance in free space. Since the substrate thickness $t_1$ is small, $\beta_1$$t_1$ can be treated as 0. Hence, the impedance of the substrate can be ignored.

Similarly, the transfer matrix of the latter parts can be expressed as

\begin{equation}
\left[\begin{array}{l}
E_3 \\
H_3
\end{array}\right]=\mathbf{M}_0\left[\begin{array}{l}
E_4 \\
H_4
\end{array}\right]=\left[\begin{array}{ll}
1 & 0 \\
\frac{1}{Z_r} & 1
\end{array}\right]\left[\begin{array}{l}
E_4 \\
H_4
\end{array}\right]
\end{equation}

\begin{equation}
\left[\begin{array}{l}
E_4 \\
H_4
\end{array}\right]=\mathbf{M}_2\left[\begin{array}{l}
E_5 \\
H_5
\end{array}\right]=\left[\begin{array}{cc}
\cos \beta h_2 & j Z_0 \sin \beta h_2 \\
j \frac{1}{Z_0} \sin \beta h_2 & \cos \beta h_2
\end{array}\right]\left[\begin{array}{l}
E_5 \\
H_5
\end{array}\right]
\end{equation}

\begin{equation}
\left[\begin{array}{l}
E_5 \\
H_5
\end{array}\right]=\mathbf{M}_b\left[\begin{array}{l}
E_6 \\
H_6
\end{array}\right]=\left[\begin{array}{ll}
1 & 0 \\
\frac{1}{z_b} & 1
\end{array}\right]\left[\begin{array}{l}
E_6 \\
H_6
\end{array}\right]
\end{equation}
where $\mathrm{Z}_r=j \omega L_0+\frac{1}{j \omega C_0}$, $\mathrm{Z}_b=j \omega L_2+\frac{1}{j \omega C_2}$. Therefore, when the EM wave is incident from port 1, the total transfer matrix can be expressed as

\begin{equation}
\begin{gathered}
\mathrm{\mathbf{T}}_{\text {total }}=\left[\begin{array}{ll}
A & B \\
C & D
\end{array}\right]=\mathbf{M}_f \mathbf{M}_1 \mathbf{M}_0 \mathbf{M}_2 \mathbf{M}_b= 
\\
{\left[\begin{array}{ll}
1 & 0 \\
\frac{1}{Z_f} & 1
\end{array}\right]\left[\begin{array}{cc}
\cos \beta h_1 & j Z_0 \sin \beta h_1 \\
j \frac{1}{Z_0} \sin \beta h_1 & \cos \beta h_1
\end{array}\right]\left[\begin{array}{cc}
1 & 0 \\
\frac{1}{z_r} & 1
\end{array}\right]}
{\left[\begin{array}{cc}
\cos \beta h_2 & j Z_0 \sin \beta h_2 \\
j \frac{1}{z_0} \sin \beta h_2 & \cos \beta h_2
\end{array}\right]\left[\begin{array}{cc}
1 & 0 \\
\frac{1}{z_b} & 1
\end{array}\right]}
\end{gathered}
\end{equation}

At this moment, we can calculate the individual parameters of the scattering matrix based on the circuit parameters of the ECM. Since the proposed metamaterial systems are reciprocal, $\operatorname{det}(T_{total})=1$ and the transmission coefficients of scattering matrix are equal, $T_f=T_b$ \cite{RN43,RN303}. We use $T$ to denote the transmission coefficient. In order to observe the complete non-Hermitian parameter space and find out the position of EP in the parameter space, we need to calculate the distribution of the eigenvalues of the metamaterial system in the space. Furthermore, we can design some practical functional devices based on metamaterial by constructing EP. According to the above analysis, the eigenvalues of the scattering matrix $\mathbf{S}_m$ can be written as

\begin{equation}
\lambda_{\pm}=T \pm \sqrt{R_f R_{\mathrm{b}}}
\end{equation}

These two eigenvalues and the corresponding eigenstates can be coalesced and lead to EP. From Equation (15), it can be concluded that the 3D-MM is $PT$ symmetric when $R_f R_b>0$. When $R_f R_b<0$, the eigenvalues form a pair of conjugate complexes, and the symmetry is broken. While when $R_f R_b=0$, the EP appears. This means that when $R_f$  or $R_b$ is 0, the phase transition from $PT$ symmetry to symmetry breaking appears. At this point, two eigenvalues coalesce, which indicates the unidirectional reflectionless appears at the EP. When the eigenvalue is 0, it means that the transmission disappears and the EP appears ($R_f R_b=0$). At this time, it can be stated that near-perfect absorption is achieved near the EP of the whole structure.

\begin{figure*}[!ht]
  \centering\includegraphics[width=6in]{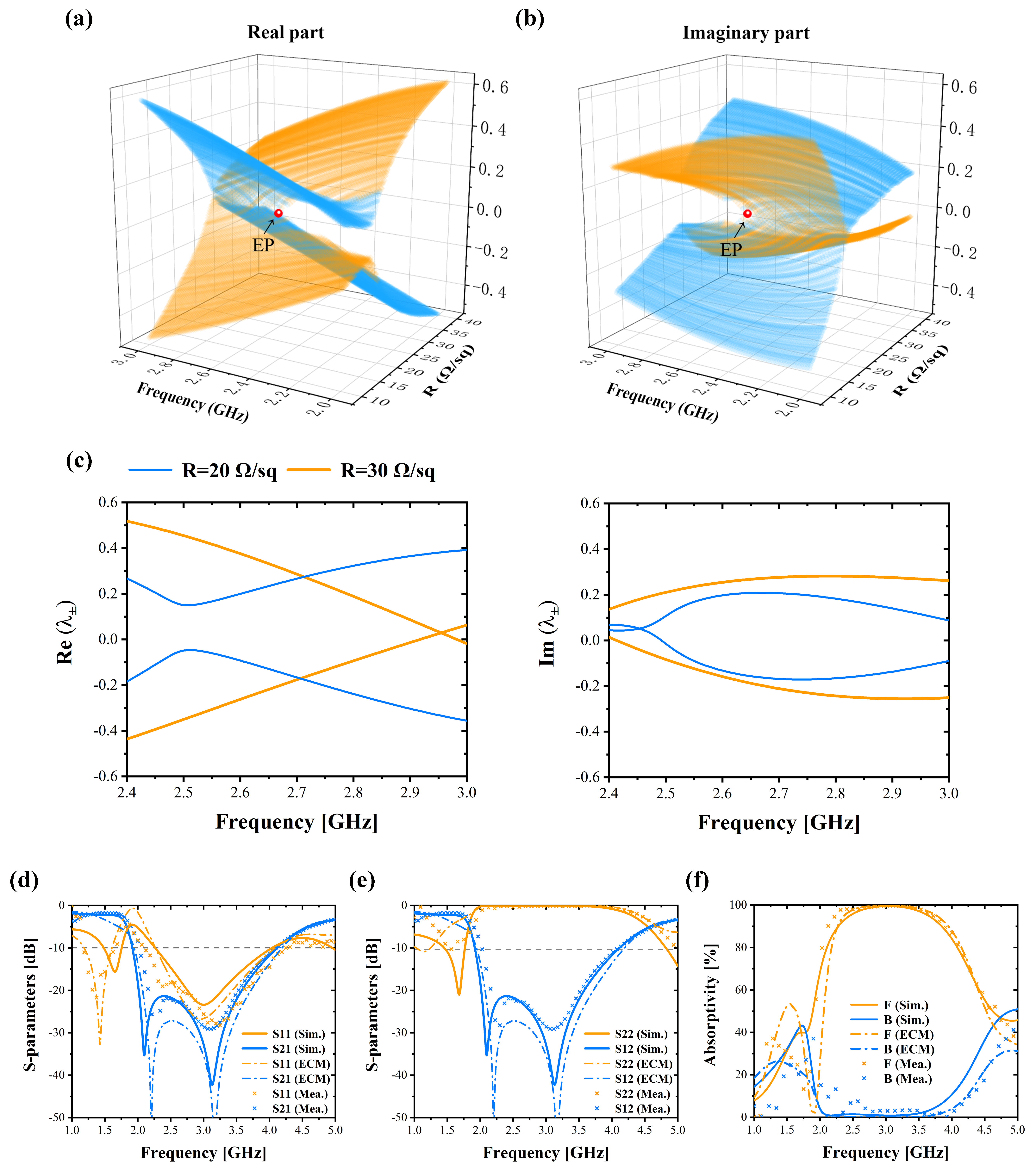}
 \caption{Eigenvalues near the EP on the self-intersecting Riemann surfaces: (a) Real part and (b) imaginary part of eigenvalues in [Frequency, $R$] parameter space. The orange and blue parts correspond to the two eigenvalues. The EP is indicated by red dots. (c) Real part and imaginary part of the scattering matrix eigenvalues ($\lambda_{\pm}$) as a function of frequency in the different sheet resistance $R$. Reflection and transmission coefficients of the proposed designs for (d) forward incidence and (e) backward incidence. (f) The absorptivity of the proposed design for forward incidence and backward incidence. The solid lines, chain lines, and forks depict the simulated, analytical calculated (ECM), and experimental results, respectively.}
  \label{Fig3}
\end{figure*}

\section*{Results and discussion}

According to the above-described analysis, the asymmetric microwave 3D-MM proposed Figure 1 is used to verify to prove the concept. In the design of the presented 3D-MM, we first set $p1$=43.4 mm, $p2$=30 mm, $h_1$=20.4 mm, $h_2$=20.4 mm, $w_1$=2.6 mm, $w_2$=2.6 mm, $c_1$=1 mm, $c_2$=1 mm, $c_3$=1 mm, $L_1$=25 mm, $L_2$=28 mm. At this point, we can verify the existence of EP by examining the eigenvalue surface of the metamaterial system. The parameter space is composed of two parameters: the frequency and loss factor of the structure. The frequency can be tuned by the incident wave, and the loss factor can also vary independently by changing the composition of graphene ink.

Figure 3(a) and Figure 3(b) show the eigenvalues near the EP on the self-intersecting Riemann surface. As can be observed from the figure, the eigenvalue surface for a given system parameter forms a self-intersecting Riemann surface with EP located at the intersection point. Figure 3(c) is the real part and imaginary part of eigenvalues ($\lambda_{\pm}$) as a function of frequency in the different sheet resistance $R$. When the $R=20$ $\Omega/sq$, the imaginary part of the eigenvalues intersects at 2.45 GHz and the real part repels. In contrast, when the $R=30$ $\Omega/sq$, the real part of the eigenvalue intersects at 2.92 GHz and the imaginary part repels. We observe that the real (imaginary) part of the eigenvalues changes from from repelling (intersecting) to intersecting (repelling) as the sheet resistance $R$ is increased. This phenomenon provides direct evidence that the EP is located within the parameter ranges $20 \leq R_{\mathrm{EP}} \leq 30$ $\Omega/sq$ and $2.45 \leq f_{\mathrm{EP}} \leq 2.92$ GHz.

Figure 3(d) and Figure 3(e) show the simulated, analytically calculated and experimental S-parameters. It can be seen from the simulated results that when the EM wave is incident from the forward direction, the $S_{11}$ and $S_{21}$ are below -10 dB in the range of 2.27-4.03 GHz (the ranges are 2.27-4.09 GHz and 2.10-4.10 GHz in calculation and experiment, respectively). When the EM wave is incident from the backward direction, the total reflection is realized. Moreover, the designed 3D-MM provides specific transmission windows in the low and high frequency bands. For the MA, when the EM wave is incident from the forward direction, the absorptivity usually is calculated according to the formula, $A_F=1-\left|S_{11}\right|^2-\left|S_{21}\right|^2$. Similarly, when the EM wave is incident from the backward direction, the absorptivity is expressed as, $A_B=1-\left|S_{22}\right|^2-\left|S_{12}\right|^2$ \cite{RN43,RN224}. As can be shown in Figure 3(f), the extremely asymmetric absorption and reflection can be observed in the range of 2.79-3.21 GHz ($absorptivity \ge 99\%$ for forward incidence and $absorptivity \le 1\%$ for backward incidence). According to the above theoretical model, the criterion for the appearance of EP is that $R_fR_b$ is 0. Here, it can be observed from Figure 3(d) and Figure 3(e) that $R_f$ is almost 0, $T$ is approximately equal to 0 ($R_f \approx T \approx 0$) and $R_b$ is also approximately 1 in the range. Therefore, extremely asymmetric absorption and reflection occurs near the EP. Similarly, the calculation results achieve EP-induced extremely asymmetric absorption and reflection in the range of 2.62-3.41 GHz. In this case, the values of optimal circuit parameters are: L$_{0}$=8.7 nH, C$_{0}$=0.29 pF, L$_{1}$=5.15 nH, C$_{1}$=0.09 pF, L$_{2}$=16.8 nH, C$_{2}$=0.31 pF, R$_{1}$=310 $\Omega$. The proposed system has high sensitivity at EP and it is susceptible to fabrication errors and experimental environment. The experimental results did not achieve perfect extremely asymmetric absorption and reflection, but the trends are in general agreement with simulated results and calculated results.

\begin{figure*}[!ht]
  \centering\includegraphics[width=6in]{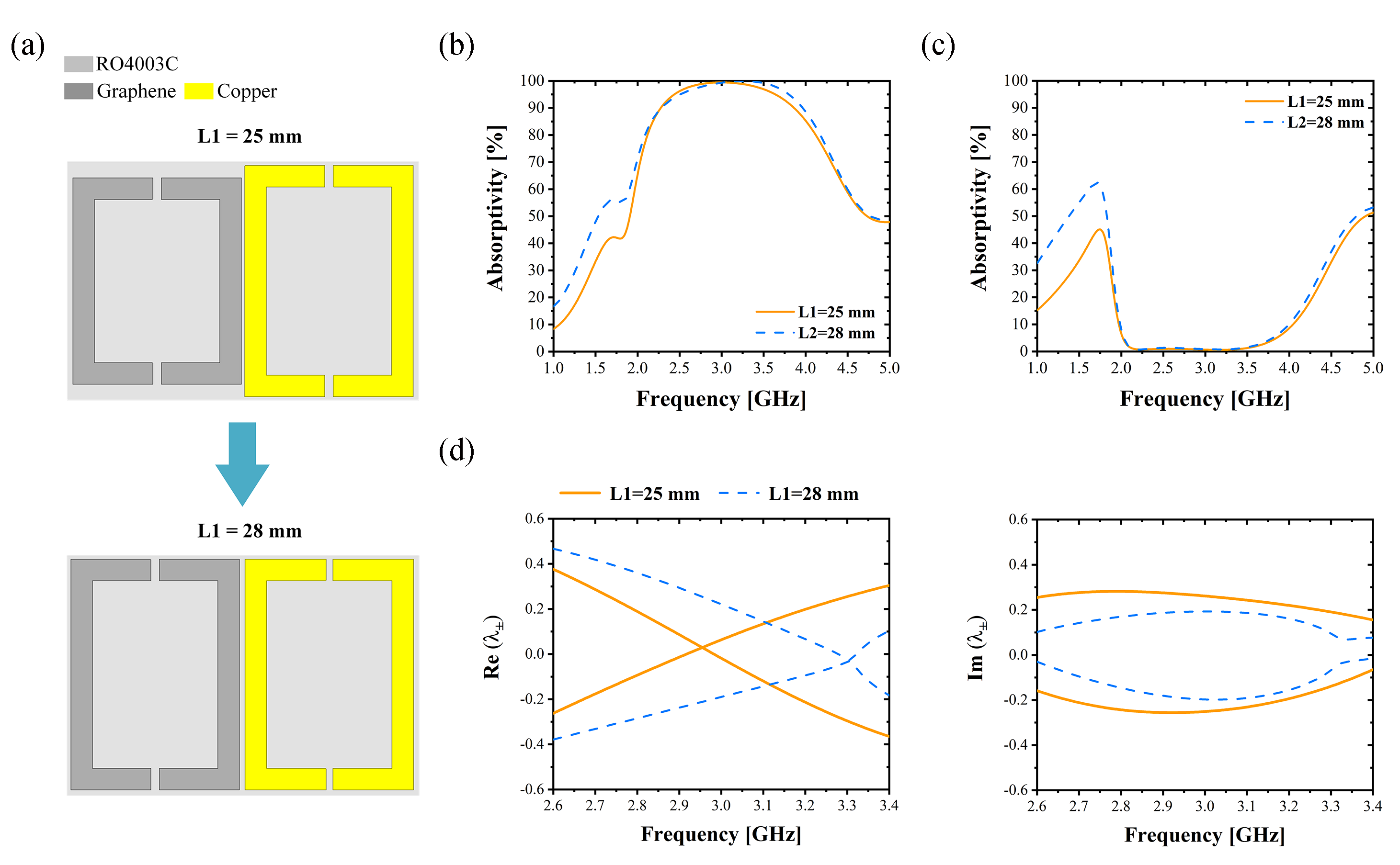}
 \caption{(a) Schematic diagram of the change of the coupling resonance length. (b) absorptivity for forward incidence; (c) absorptivity for backward incidence. (d) Real part and imaginary part of the scattering matrix eigenvalues ($\lambda_{\pm}$) as a function of frequency when the coupling length changes.}
  \label{Fig4}
\end{figure*}

Moreover, it has been shown in previous research that the absorption and reflection of waves are manipulated by the coupling resonance between SRRs \cite{RN43}. Therefore, adjusting the coupling part of the metal strip can also adjust the absorption and reflection performance of the 3D-MM. As shown in Figure 4(b), when the smaller SRR is increased to the same size as the larger SRR (i.e. L1=28 mm), the absorption band has been significantly broadened. It achieves broadband absorption in the range of 2.28 - 3.97 GHz, and the absorptivity is greater than 90\%. As shown in Figure 4(c), when the EM wave is incident in the opposite direction, the reflection band remains almost unchanged. Figure 4(b) and Figure 4(c) illustrate that extremely asymmetric absorption and reflection are achieved at this moment. Figure 4(d) is the real part and imaginary part of eigenvalues ($\lambda_{\pm}$) as a function of frequency when the coupling length changes. It can be observed from the figure that the coalescing point of the real part is shifted to 3.3 GHz and the imaginary part repels. With the change of structure, the parameter space changes and the location of the EP moves. Comparatively, when loaded with the sheet resistance R=30 $\Omega/sq$, it is closer to the EP at L=28 mm than at L=25 mm. As a result, a more significant phenomenon can be observed.

These two proof-of-concept designs indicate that tuning the intrinsic loss and geometric parameters of the system is an efficient way to achieve EP-induced asymmetric absorption and reflection. Therefore, by appropriately tuning the geometric parameter and the intrinsic loss of the system, some advantageous properties can be achieved. 

\section*{Experiment and test}

\begin{table}[ht]\footnotesize
 \centering
\begin{threeparttable}
\renewcommand{\arraystretch}{2}
 \caption{Materials used in graphene ink}

    \begin{tabular}[htbp]{@{}m{3.5cm}<{\centering}m{3.5cm}<{\centering}m{3.5cm}<{\centering}m{3.5 cm}@{}}
    \hline
    Component & Material & Mass(g) \\
    \hline

    \multirow{2}{*}{Conductive filler}
          & GEAS{$^1$} (weight 5\%)& 51.36 \\
                       & carbon black   & 1.95\\\hline
     \multirow{2}{*}{ Solvent}
     & deionized water  & 5.12  \\
                   & propylene glycol   & 4.86\\\hline

     \multirow{5}{*}{Additive agent}
    & XR 500{$^2$}   & 0.28\\
        & CMCC{$^3$} & 0.13 \\ 
       & 9300{$^4$} & 0.29 \\ 
     &   Silok 4600{$^5$} & 0.26 \\ 
     &   Acrylic resin{$^6$} & 4.05 \\ 
     \hline

     
    \hline
  \end{tabular}

\begin{tablenotes}
\item[1]GEAS graphene electric aqueous slurry.
\item[2]XR 500: cross-linking agent.
\item[3]CMCC: sodium carboxymethyl cellulose (thickening agent).
\item[4]9300: sodium polyacrylate ACUMER 9300 (dispersant agent).
\item[5]Silok 4600: defoaming agent.
\item[6]Acrylic resin: binding agent.
\end{tablenotes}
\end{threeparttable}
  
\end{table}

\begin{figure*}[!ht]
  \centering\includegraphics[width=6in]{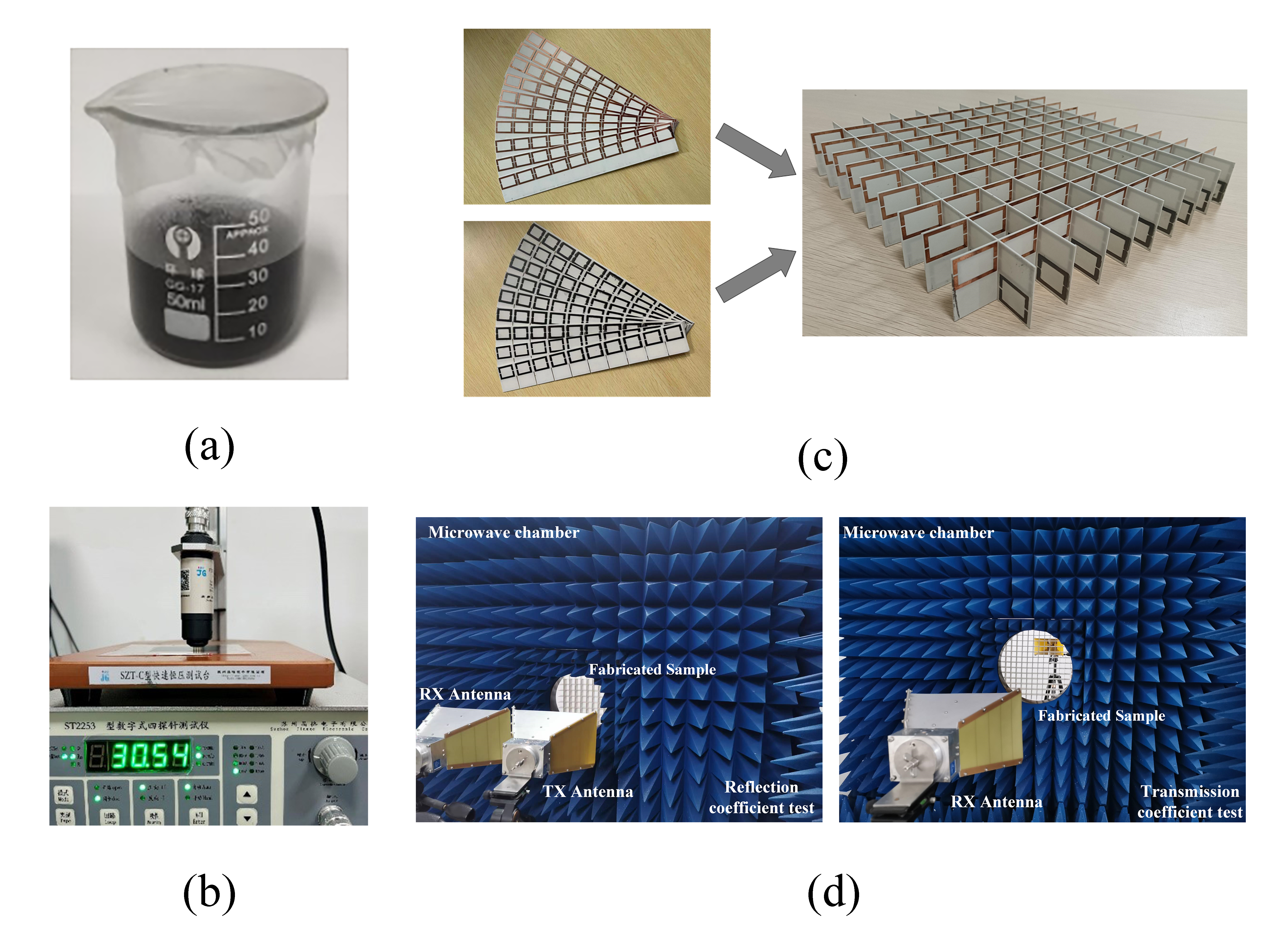}
 \caption{(a) The prepared graphene ink. (b) Sheet resistance measurement of the prepared graphene ink. (c) The printed graphene-based resistive film substrate and prototype of the manufactured 3D-MM. (d) Measurement environment: reflection coefficient test and transmission coefficient test.}
  \label{fig:boat1}
\end{figure*}

In this paper, we use graphene-based resistive ink and screen-printing technology to realise the lossy metamaterial systems~\cite{RN45,RN46}. The graphene ink is composed of conductive fillers, solvents and additive agents, and the specific ingredients are shown in Table 1. It is noted that the graphene electric aqueous slurry (GEAS) is used as the main conductive filler due to its high electrical conductivity. Besides, since the lamellar distribution of graphene, black carbon (BC) is adopted to form conductive pathways between the graphene layers. While excessive BC is adverse to reducing the conductivity of the graphene ink due to its lower electrical conductivity compared with graphene. Some other additive agents such as cross-linking agent, thickening agent and dispersant agent are used to adjust the properties and state of the graphene ink.

The specific fabrication process could be divided into three steps. Firstly, weigh the components in Table 1 and pre-mix them in a magnetic mixer. Then, defoam and re-mixed the conductive ink in a planetary mixer/deaerator (MAZERUSTAR KK300SSE) at a speed of 2000 rpm for 4 minutes. Finally, disperse the graphene ink by an ultrasonic cleaner (031ST) for 30 minutes to ensure that the BC and graphene flakes are evenly dispersed without aggregation. After the above three steps, the prepared graphene ink could be printed on Printed circuit boards (PCB) evenly by screen-printing technology as shown in Figure 5(a). The sheet resistance of the fabricated graphene ink is measured by a 4-probe tester as shown in Figure 5(b). It shows that the sheet resistance of prepared graphene ink is 30.54 $\Omega/sq$ and fully meets the simulation and measurement requirements.

To better verify the concept, the prototype of the designed 3D-MM was fabricated and measured. Since the printed graphene ink is easy to permeate other structures, the graphene-based resistive film is printed on the back of the substrate. The prepared ink was printed on a 0.813 mm thick Rogers RO4003C substrate by screen printing processing. Subsequently, assemble the grooved substrates and the assembled 3D-MM are shown in Figure 5(c). The fabricated prototype includes 10 × 10 unit cells and the total dimension is 300 × 300 × 43.4 $mm^3$. As shown in Figure 5(d), the designed 3D-MM prototype was measured in a microwave chamber. A pair of horn antennas operating from 1 to 18 GHz is connected to a vector network analyzer (Agilent E8362B) is used to measure the reflection and transmission coefficients of the prototype. The experimental results are shown in Figure 3(d), Figure 3(e) and Figure 3(f). The deviations between the simulated and experimental results may originate from the interference of experimental environmental noise, the non-uniformity of the graphene ink printing, the fabrication tolerance of the sample, the instability of the relative dielectric constant of the substrate, and other fabrication errors.

\section*{Conclusion}
 
In this paper, a 3D-MM system modeled near its EP is proposed, which achieves extremely asymmetric absorption and reflection. We combine the equivalent circuit model with the Hamiltonian quantum physical model to construct the non-Hermitian transmission matrix of the designed structure. Meanwhile, metamaterial systems with EP response can be generated by tuning the structure and circuit parameters of the ECM, which contributes in the design of asymmetric metamaterial absorbers. The designed metamaterial system consists of a hollow vertical periodic crossed mesh array without the metal backing plate. By introducing a graphene-based resistive film to provide localized losses, the 3D-MM realizes the extremely asymmetric absorption and reflection near the EP. Moreover, the designed metamaterial system can achieve $90\%$ broadband absorption near EP at 2.28-3.85 GHz from the forward incidence. When the EM wave is incident from the backward direction, total reflection is realized. Simultaneously, out of the absorbing band, there are low insertion loss transmission windows. This phenomenon is verified both theoretically and experimentally. This paper demonstrates that the conventional equivalent circuit theory can be used to construct non-Hermitian systems. Our results may help to further extend their potential applications in novel EM functional devices, for example in the design of perfect MAs, sensors, and other novel devices, etc.

\section*{Data availability statement}
The data that support the findings of this study are available from the corresponding author upon reasonable request.

\section*{Acknowledgements}

This work was supported by the National Natural Science Foundation of China (U20B2043, 62001095, 11804087); Financially supported by self-determined research funds of CCNU from the colleges’basic research and operation of MOE” (CCNU19TS073, CCNU20GF004, CCNU23CG016); Guangxi Key Laboratory of Wireless Wideband Communication and Signal Processing, Guilin University of Electronic Technology (GXKL06190202); Open fund of China Ship Development and Design Centre (XM0120190196); Science and Technology Research Project of Hubei Education Department (D20222903); Hongque Innovation Center (HQ202104001); Hubei University (X202210512039, 030-017643); and Hubei Province Innovation Special Fund (2023BAB061).

\section*{Author contributions}

Yanjie Wu: Conceptualization; Methodology; Validation; Calculation; Produced all the figures and wrote the paper draft. Ding Zhang: Resources; Writing-Original Draft. Qiuyu Li and Xintong Shi: Investigation; Writing-Original Draft. Jie Xiong, Haoquan Hu, Jing Tian and Bian Wu: Funding acquisition; Writing-Review \& Editing. Hai Lin: Project administration; Funding acquisition; Writing-Review \& Editing. Y. Liu: Supervision; Funding acquisition; Writing-Review \& Editing. All of the authors discussed the results and commented on the manuscript. All authors read and approved the final manuscript.

\section*{Competing interests}

The authors declare no competing interests.


\bibliography{sample}

\begin{thebibliography}{10}
\urlstyle{rm}
\expandafter\ifx\csname url\endcsname\relax
  \def\url#1{\texttt{#1}}\fi
\expandafter\ifx\csname urlprefix\endcsname\relax\def\urlprefix{URL }\fi
\expandafter\ifx\csname doiprefix\endcsname\relax\def\doiprefix{DOI: }\fi
\providecommand{\bibinfo}[2]{#2}
\providecommand{\eprint}[2][]{\url{#2}}

\bibitem{RN102}
\bibinfo{author}{Mei, Z.}, \bibinfo{author}{Zhang, L.} \& \bibinfo{author}{Cui, T.}
\newblock \bibinfo{journal}{\bibinfo{title}{Recent advances on metamaterials}}.
\newblock {\emph{\JournalTitle{Science \& Technology Review}}} \textbf{\bibinfo{volume}{34}}, \bibinfo{pages}{27--39}, \doiprefix\url{10.3981/j.issn.1000-7857.2016.18.002} (\bibinfo{year}{2016}).

\bibitem{RN273}
\bibinfo{author}{Wang, Y.-Z.}, \bibinfo{author}{Xu, H.-X.}, \bibinfo{author}{Wang, C.-H.}, \bibinfo{author}{Wang, M.-Z.} \& \bibinfo{author}{Wang, S.-J.}
\newblock \bibinfo{journal}{\bibinfo{title}{Research progress of electromagnetic metamaterial absorbers}}.
\newblock {\emph{\JournalTitle{Acta Physica Sinica}}} \textbf{\bibinfo{volume}{69}}, \bibinfo{pages}{134101}, \doiprefix\url{10.7498/aps.69.20200355} (\bibinfo{year}{2020}).

\bibitem{RN101}
\bibinfo{author}{Zheludev, N.~I.} \& \bibinfo{author}{Kivshar, Y.~S.}
\newblock \bibinfo{journal}{\bibinfo{title}{From metamaterials to metadevices}}.
\newblock {\emph{\JournalTitle{Nat Mater}}} \textbf{\bibinfo{volume}{11}}, \bibinfo{pages}{917--24}, \doiprefix\url{10.1038/nmat3431} (\bibinfo{year}{2012}).

\bibitem{RN103}
\bibinfo{author}{Miri, M.~A.} \& \bibinfo{author}{Alu, A.}
\newblock \bibinfo{journal}{\bibinfo{title}{Exceptional points in optics and photonics}}.
\newblock {\emph{\JournalTitle{Science}}} \textbf{\bibinfo{volume}{363}}, \bibinfo{pages}{aar7709}, \doiprefix\url{10.1126/science.aar7709} (\bibinfo{year}{2019}).

\bibitem{RN955}
\bibinfo{author}{Boltasseva, A.} \& \bibinfo{author}{Atwater, H.~A.}
\newblock \bibinfo{journal}{\bibinfo{title}{Low-loss plasmonic metamaterials}}.
\newblock {\emph{\JournalTitle{Science}}} \textbf{\bibinfo{volume}{331}}, \bibinfo{pages}{290--291}, \doiprefix\url{10.1126/science.1198258} (\bibinfo{year}{2011}).

\bibitem{RN104}
\bibinfo{author}{dos Santos, J. F.~G.}, \bibinfo{author}{Luiz, F.~S.}, \bibinfo{author}{Duarte, O.~S.} \& \bibinfo{author}{Moussa, M. H.~Y.}
\newblock \bibinfo{journal}{\bibinfo{title}{Non-hermitian noncommutative quantum mechanics}}.
\newblock {\emph{\JournalTitle{The European Physical Journal Plus}}} \textbf{\bibinfo{volume}{134}}, \bibinfo{pages}{332}, \doiprefix\url{10.1140/epjp/i2019-12738-3} (\bibinfo{year}{2019}).

\bibitem{RN105}
\bibinfo{author}{Yang, M.}, \bibinfo{author}{Ye, Z.}, \bibinfo{author}{Farhat, M.} \& \bibinfo{author}{Chen, P.-Y.}
\newblock \bibinfo{journal}{\bibinfo{title}{Cascaded pt-symmetric artificial sheets: multimodal manipulation of self-dual emitter-absorber singularities, and unidirectional and bidirectional reflectionless transparencies}}.
\newblock {\emph{\JournalTitle{Journal of Physics D: Applied Physics}}} \textbf{\bibinfo{volume}{55}}, \bibinfo{pages}{085301}, \doiprefix\url{10.1088/1361-6463/ac3300} (\bibinfo{year}{2021}).

\bibitem{RN107}
\bibinfo{author}{Huang, Y.}, \bibinfo{author}{Veronis, G.} \& \bibinfo{author}{Min, C.}
\newblock \bibinfo{journal}{\bibinfo{title}{Unidirectional reflectionless propagation in plasmonic waveguide-cavity systems at exceptional points}}.
\newblock {\emph{\JournalTitle{Optics Express}}} \textbf{\bibinfo{volume}{23}}, \bibinfo{pages}{29882--29895}, \doiprefix\url{10.1364/OE.23.029882} (\bibinfo{year}{2015}).

\bibitem{RN109}
\bibinfo{author}{Wang, X.}, \bibinfo{author}{Fang, X.}, \bibinfo{author}{Mao, D.}, \bibinfo{author}{Jing, Y.} \& \bibinfo{author}{Li, Y.}
\newblock \bibinfo{journal}{\bibinfo{title}{Extremely asymmetrical acoustic metasurface mirror at the exceptional point}}.
\newblock {\emph{\JournalTitle{Physical Review Letters}}} \textbf{\bibinfo{volume}{123}}, \bibinfo{pages}{214302}, \doiprefix\url{10.1103/PhysRevLett.123.214302} (\bibinfo{year}{2019}).

\bibitem{RN111}
\bibinfo{author}{Hodaei, H.}, \bibinfo{author}{Miri, M.-A.}, \bibinfo{author}{Heinrich, M.}, \bibinfo{author}{Christodoulides, D.~N.} \& \bibinfo{author}{Khajavikhan, M.}
\newblock \bibinfo{journal}{\bibinfo{title}{Parity-time-symmetric microring lasers}}.
\newblock {\emph{\JournalTitle{Science}}} \textbf{\bibinfo{volume}{346}}, \bibinfo{pages}{975--978}, \doiprefix\url{doi:10.1126/science.1258480} (\bibinfo{year}{2014}).

\bibitem{RN113}
\bibinfo{author}{Bender, C.~M.} \& \bibinfo{author}{Boettcher, S.}
\newblock \bibinfo{journal}{\bibinfo{title}{Real spectra in non-hermitian hamiltonians having pt symmetry}}.
\newblock {\emph{\JournalTitle{Physical Review Letters}}} \textbf{\bibinfo{volume}{80}}, \bibinfo{pages}{5243--5246}, \doiprefix\url{10.1103/PhysRevLett.80.5243} (\bibinfo{year}{1998}).

\bibitem{RN114}
\bibinfo{author}{Lee, S.-B.} \emph{et~al.}
\newblock \bibinfo{journal}{\bibinfo{title}{Observation of an exceptional point in a chaotic optical microcavity}}.
\newblock {\emph{\JournalTitle{Physical Review Letters}}} \textbf{\bibinfo{volume}{103}}, \bibinfo{pages}{134101}, \doiprefix\url{10.1103/PhysRevLett.103.134101} (\bibinfo{year}{2009}).

\bibitem{RN115}
\bibinfo{author}{Wang, X.-Y.}, \bibinfo{author}{Wang, F.-F.} \& \bibinfo{author}{Hu, X.-Y.}
\newblock \bibinfo{journal}{\bibinfo{title}{Waveguide-induced coalescence of exceptional points}}.
\newblock {\emph{\JournalTitle{Physical Review A}}} \textbf{\bibinfo{volume}{101}}, \bibinfo{pages}{053820}, \doiprefix\url{10.1103/PhysRevA.101.053820} (\bibinfo{year}{2020}).

\bibitem{RN117}
\bibinfo{author}{Mann, S.} \emph{et~al.}
\newblock \bibinfo{journal}{\bibinfo{title}{How to achieve exceptional points in coupled resonators using a gyrator or pt-symmetry, and in a time-modulated single resonator: high sensitivity to perturbations}}.
\newblock {\emph{\JournalTitle{EPJ Applied Metamaterials}}} \textbf{\bibinfo{volume}{9}}, \bibinfo{pages}{14}, \doiprefix\url{10.1051/epjam/2022006} (\bibinfo{year}{2022}).

\bibitem{RN118}
\bibinfo{author}{Ding, K.}, \bibinfo{author}{Fang, C.} \& \bibinfo{author}{Ma, G.}
\newblock \bibinfo{journal}{\bibinfo{title}{Non-hermitian topology and exceptional-point geometries}}.
\newblock {\emph{\JournalTitle{Nature Reviews Physics}}} \textbf{\bibinfo{volume}{4}}, \bibinfo{pages}{745--760}, \doiprefix\url{10.1038/s42254-022-00516-5} (\bibinfo{year}{2022}).

\bibitem{RN119}
\bibinfo{author}{An, S.} \emph{et~al.}
\newblock \bibinfo{journal}{\bibinfo{title}{Unidirectional invisibility of an acoustic multilayered medium with parity-time-symmetric impedance modulation}}.
\newblock {\emph{\JournalTitle{Journal of Applied Physics}}} \textbf{\bibinfo{volume}{129}}, \bibinfo{pages}{175106}, \doiprefix\url{10.1063/5.0039432} (\bibinfo{year}{2021}).

\bibitem{RN121}
\bibinfo{author}{Ye, Z.}, \bibinfo{author}{Yang, M.}, \bibinfo{author}{Zhu, L.} \& \bibinfo{author}{Chen, P.-Y.}
\newblock \bibinfo{journal}{\bibinfo{title}{Ptx-symmetric metasurfaces for sensing applications}}.
\newblock {\emph{\JournalTitle{Frontiers of Optoelectronics}}} \textbf{\bibinfo{volume}{14}}, \bibinfo{pages}{211--220}, \doiprefix\url{10.1007/s12200-021-1204-6} (\bibinfo{year}{2021}).

\bibitem{RN123}
\bibinfo{author}{Zhang, H.}, \bibinfo{author}{Saif, F.}, \bibinfo{author}{Jiao, Y.} \& \bibinfo{author}{Jing, H.}
\newblock \bibinfo{journal}{\bibinfo{title}{Loss-induced transparency in optomechanics}}.
\newblock {\emph{\JournalTitle{Opt Express}}} \textbf{\bibinfo{volume}{26}}, \bibinfo{pages}{25199--25210}, \doiprefix\url{10.1364/OE.26.025199} (\bibinfo{year}{2018}).

\bibitem{RN125}
\bibinfo{author}{Laha, A.}, \bibinfo{author}{Dey, S.}, \bibinfo{author}{Gandhi, H.~K.}, \bibinfo{author}{Biswas, A.} \& \bibinfo{author}{Ghosh, S.}
\newblock \bibinfo{journal}{\bibinfo{title}{Exceptional point and toward mode-selective optical isolation}}.
\newblock {\emph{\JournalTitle{ACS Photonics}}} \textbf{\bibinfo{volume}{7}}, \bibinfo{pages}{967--974}, \doiprefix\url{10.1021/acsphotonics.9b01646} (\bibinfo{year}{2020}).

\bibitem{RN127}
\bibinfo{author}{Hodaei, H.}, \bibinfo{author}{Miri, M.-A.}, \bibinfo{author}{Heinrich, M.}, \bibinfo{author}{Christodoulides, D.~N.} \& \bibinfo{author}{Khajavikhan, M.}
\newblock \bibinfo{journal}{\bibinfo{title}{Parity-time-symmetric microring lasers}}.
\newblock {\emph{\JournalTitle{Science}}} \textbf{\bibinfo{volume}{346}}, \bibinfo{pages}{975--978}, \doiprefix\url{doi:10.1126/science.1258480} (\bibinfo{year}{2014}).

\bibitem{RN110}
\bibinfo{author}{Li, D.}, \bibinfo{author}{Huang, S.}, \bibinfo{author}{Cheng, Y.} \& \bibinfo{author}{Li, Y.}
\newblock \bibinfo{journal}{\bibinfo{title}{Compact asymmetric sound absorber at the exceptional point}}.
\newblock {\emph{\JournalTitle{Science China Physics, Mechanics \& Astronomy}}} \textbf{\bibinfo{volume}{64}}, \bibinfo{pages}{244303}, \doiprefix\url{10.1007/s11433-020-1612-1} (\bibinfo{year}{2021}).

\bibitem{RN930_45}
\bibinfo{author}{Zhang, H.} \emph{et~al.}
\newblock \bibinfo{journal}{\bibinfo{title}{Topological bound state in the continuum induced unidirectional acoustic perfect absorption}}.
\newblock {\emph{\JournalTitle{Science China Physics, Mechanics \& Astronomy}}} \textbf{\bibinfo{volume}{66}}, \bibinfo{pages}{284311}, \doiprefix\url{10.1007/s11433-023-2136-y} (\bibinfo{year}{2023}).

\bibitem{RN948}
\bibinfo{author}{Yang, Y.} \emph{et~al.}
\newblock \bibinfo{journal}{\bibinfo{title}{Radiative anti-parity-time plasmonics}}.
\newblock {\emph{\JournalTitle{Nat Commun}}} \textbf{\bibinfo{volume}{13}}, \bibinfo{pages}{7678}, \doiprefix\url{10.1038/s41467-022-35447-3} (\bibinfo{year}{2022}).

\bibitem{RN957_40}
\bibinfo{author}{Zhu, X.}, \bibinfo{author}{Ramezani, H.}, \bibinfo{author}{Shi, C.}, \bibinfo{author}{Zhu, J.} \& \bibinfo{author}{Zhang, X.}
\newblock \bibinfo{journal}{\bibinfo{title}{Pt-symmetric acoustics}}.
\newblock {\emph{\JournalTitle{Physical Review X}}} \textbf{\bibinfo{volume}{4}}, \bibinfo{pages}{031042}, \doiprefix\url{10.1103/PhysRevX.4.031042} (\bibinfo{year}{2014}).

\bibitem{RN938_41}
\bibinfo{author}{Fan, H.-Y.} \& \bibinfo{author}{Luo, J.}
\newblock \bibinfo{journal}{\bibinfo{title}{Research progress of non-hermitian electromagnetic metasurfaces}}.
\newblock {\emph{\JournalTitle{Acta Physica Sinica}}} \textbf{\bibinfo{volume}{71}}, \bibinfo{pages}{247802}, \doiprefix\url{10.7498/aps.71.20221706} (\bibinfo{year}{2022}).

\bibitem{RN108}
\bibinfo{author}{Feng, L.} \emph{et~al.}
\newblock \bibinfo{journal}{\bibinfo{title}{Experimental demonstration of a unidirectional reflectionless parity-time metamaterial at optical frequencies}}.
\newblock {\emph{\JournalTitle{Nature Materials}}} \textbf{\bibinfo{volume}{12}}, \bibinfo{pages}{108--113}, \doiprefix\url{10.1038/nmat3495} (\bibinfo{year}{2013}).

\bibitem{RN129}
\bibinfo{author}{Yang, F.}, \bibinfo{author}{Hwang, A.}, \bibinfo{author}{Doiron, C.} \& \bibinfo{author}{Naik, G.~V.}
\newblock \bibinfo{journal}{\bibinfo{title}{Non-hermitian metasurfaces for the best of plasmonics and dielectrics}}.
\newblock {\emph{\JournalTitle{Optical Materials Express}}} \textbf{\bibinfo{volume}{11}}, \bibinfo{pages}{2326}, \doiprefix\url{10.1364/ome.428469} (\bibinfo{year}{2021}).

\bibitem{RN130}
\bibinfo{author}{Park, S.~H.} \emph{et~al.}
\newblock \bibinfo{journal}{\bibinfo{title}{Observation of an exceptional point in a non-hermitian metasurface}}.
\newblock {\emph{\JournalTitle{Nanophotonics}}} \textbf{\bibinfo{volume}{9}}, \bibinfo{pages}{1031--1039}, \doiprefix\url{10.1515/nanoph-2019-0489} (\bibinfo{year}{2020}).

\bibitem{RN135}
\bibinfo{author}{Huang, Y.}, \bibinfo{author}{Veronis, G.} \& \bibinfo{author}{Min, C.}
\newblock \bibinfo{journal}{\bibinfo{title}{Unidirectional reflectionless propagation in plasmonic waveguide-cavity systems at exceptional points}}.
\newblock {\emph{\JournalTitle{Optics Express}}} \textbf{\bibinfo{volume}{23}}, \bibinfo{pages}{29882--29895}, \doiprefix\url{10.1364/OE.23.029882} (\bibinfo{year}{2015}).

\bibitem{RN919_42}
\bibinfo{author}{Kang, M.}, \bibinfo{author}{Zhang, T.}, \bibinfo{author}{Zhao, B.}, \bibinfo{author}{Sun, L.} \& \bibinfo{author}{Chen, J.}
\newblock \bibinfo{journal}{\bibinfo{title}{Chirality of exceptional points in bianisotropic metasurfaces}}.
\newblock {\emph{\JournalTitle{Opt Express}}} \textbf{\bibinfo{volume}{29}}, \bibinfo{pages}{11582--11590}, \doiprefix\url{10.1364/OE.419511} (\bibinfo{year}{2021}).

\bibitem{RN136}
\bibinfo{author}{Dong, S.} \emph{et~al.}
\newblock \bibinfo{journal}{\bibinfo{title}{Loss-assisted metasurface at an exceptional point}}.
\newblock {\emph{\JournalTitle{ACS Photonics}}} \textbf{\bibinfo{volume}{7}}, \bibinfo{pages}{3321--3327}, \doiprefix\url{10.1021/acsphotonics.0c01440} (\bibinfo{year}{2020}).

\bibitem{RN953_44}
\bibinfo{author}{Li, M.}, \bibinfo{author}{Wang, Z.}, \bibinfo{author}{Yin, W.-Y.}, \bibinfo{author}{Li, E.-P.} \& \bibinfo{author}{Chen, H.}
\newblock \bibinfo{journal}{\bibinfo{title}{Controlling asymmetric retroreflection of metasurfaces via localized loss engineering}}.
\newblock {\emph{\JournalTitle{IEEE Transactions on Antennas and Propagation}}} \textbf{\bibinfo{volume}{70}}, \bibinfo{pages}{11858--11866}, \doiprefix\url{10.1109/tap.2022.3209282} (\bibinfo{year}{2022}).

\bibitem{RN922_46}
\bibinfo{author}{Alaee, R.} \emph{et~al.}
\newblock \bibinfo{journal}{\bibinfo{title}{Magnetoelectric coupling in nonidentical plasmonic nanoparticles: Theory and applications}}.
\newblock {\emph{\JournalTitle{Physical Review B}}} \textbf{\bibinfo{volume}{91}}, \bibinfo{pages}{115119}, \doiprefix\url{10.1103/PhysRevB.91.115119} (\bibinfo{year}{2015}).

\bibitem{RN923_47}
\bibinfo{author}{Kriegler, C.~E.}, \bibinfo{author}{Rill, M.~S.}, \bibinfo{author}{Linden, S.} \& \bibinfo{author}{Wegener, M.}
\newblock \bibinfo{journal}{\bibinfo{title}{Bianisotropic photonic metamaterials}}.
\newblock {\emph{\JournalTitle{IEEE Journal of Selected Topics in Quantum Electronics}}} \textbf{\bibinfo{volume}{16}}, \bibinfo{pages}{367--375}, \doiprefix\url{10.1109/jstqe.2009.2020809} (\bibinfo{year}{2010}).

\bibitem{RN155_48}
\bibinfo{author}{Zhixin, Y.}, \bibinfo{author}{Shaoqiu, X.}, \bibinfo{author}{Li, Y.} \& \bibinfo{author}{Wang, B.-Z.}
\newblock \bibinfo{journal}{\bibinfo{title}{On the design of wideband absorber based on multilayer and multiresonant fss array}}.
\newblock {\emph{\JournalTitle{IEEE Antennas and Wireless Propagation Letters}}} \textbf{\bibinfo{volume}{20}}, \bibinfo{pages}{284--288}, \doiprefix\url{10.1109/lawp.2020.3046010} (\bibinfo{year}{2021}).

\bibitem{RN836_49}
\bibinfo{author}{Zhang, D.} \emph{et~al.}
\newblock \bibinfo{journal}{\bibinfo{title}{Ultra-wideband flexible radar-infrared bi-stealth absorber based on a patterned graphene}}.
\newblock {\emph{\JournalTitle{Opt Express}}} \textbf{\bibinfo{volume}{31}}, \bibinfo{pages}{1969--1981}, \doiprefix\url{10.1364/OE.476639} (\bibinfo{year}{2023}).

\bibitem{RN106}
\bibinfo{author}{Sakhdari, M.}, \bibinfo{author}{Farhat, M.} \& \bibinfo{author}{Chen, P.-Y.}
\newblock \bibinfo{journal}{\bibinfo{title}{Pt-symmetric metasurfaces: wave manipulation and sensing using singular points}}.
\newblock {\emph{\JournalTitle{New Journal of Physics}}} \textbf{\bibinfo{volume}{19}}, \bibinfo{pages}{065002}, \doiprefix\url{10.1088/1367-2630/aa6bb9} (\bibinfo{year}{2017}).

\bibitem{RN966_50}
\bibinfo{author}{Sakotic, Z.}, \bibinfo{author}{Krasnok, A.}, \bibinfo{author}{Alú, A.} \& \bibinfo{author}{Jankovic, N.}
\newblock \bibinfo{journal}{\bibinfo{title}{Topological scattering singularities and embedded eigenstates for polarization control and sensing applications}}.
\newblock {\emph{\JournalTitle{Photonics Research}}} \textbf{\bibinfo{volume}{9}}, \bibinfo{pages}{1310}, \doiprefix\url{10.1364/prj.424247} (\bibinfo{year}{2021}).

\bibitem{RN936}
\bibinfo{author}{Sakotic, Z.} \emph{et~al.}
\newblock \bibinfo{journal}{\bibinfo{title}{Non‐hermitian control of topological scattering singularities emerging from bound states in the continuum}}.
\newblock {\emph{\JournalTitle{Laser \& Photonics Reviews}}} \textbf{\bibinfo{volume}{17}}, \bibinfo{pages}{2200308}, \doiprefix\url{10.1002/lpor.202200308} (\bibinfo{year}{2023}).

\bibitem{RN43}
\bibinfo{author}{Lin, H.} \emph{et~al.}
\newblock \bibinfo{journal}{\bibinfo{title}{Dual-polarized bidirectional three-dimensional metamaterial absorber with transmission windows}}.
\newblock {\emph{\JournalTitle{Optics Express}}} \textbf{\bibinfo{volume}{29}}, \bibinfo{pages}{40770--40780}, \doiprefix\url{10.1364/oe.446143} (\bibinfo{year}{2021}).

\bibitem{RN925}
\bibinfo{author}{Ramezani, H.}, \bibinfo{author}{Wang, Y.}, \bibinfo{author}{Yablonovitch, E.} \& \bibinfo{author}{Zhang, X.}
\newblock \bibinfo{journal}{\bibinfo{title}{Unidirectional perfect absorber}}.
\newblock {\emph{\JournalTitle{IEEE Journal of Selected Topics in Quantum Electronics}}} \textbf{\bibinfo{volume}{22}}, \bibinfo{pages}{115--120}, \doiprefix\url{10.1109/jstqe.2016.2545644} (\bibinfo{year}{2016}).

\bibitem{RN298}
\bibinfo{author}{Landy, N.~I.}, \bibinfo{author}{Sajuyigbe, S.}, \bibinfo{author}{Mock, J.~J.}, \bibinfo{author}{Smith, D.~R.} \& \bibinfo{author}{Padilla, W.~J.}
\newblock \bibinfo{journal}{\bibinfo{title}{Perfect metamaterial absorber}}.
\newblock {\emph{\JournalTitle{Phys Rev Lett}}} \textbf{\bibinfo{volume}{100}}, \bibinfo{pages}{207402}, \doiprefix\url{10.1103/PhysRevLett.100.207402} (\bibinfo{year}{2008}).

\bibitem{RN390}
\bibinfo{author}{He, F.} \emph{et~al.}
\newblock \bibinfo{journal}{\bibinfo{title}{Broadband frequency selective surface absorber with dual-section step-impedance matching for oblique incidence applications}}.
\newblock {\emph{\JournalTitle{IEEE Transactions on Antennas and Propagation}}} \textbf{\bibinfo{volume}{69}}, \bibinfo{pages}{7647--7657}, \doiprefix\url{10.1109/tap.2021.3070065} (\bibinfo{year}{2021}).

\bibitem{RN826}
\bibinfo{author}{Deane, J. H.~B.} \& \bibinfo{author}{Johnstone, G.~G.}
\newblock \bibinfo{journal}{\bibinfo{title}{Brief communication matrix equations for the relations between two-port parameters}}.
\newblock {\emph{\JournalTitle{International Journal of Electronics}}} \textbf{\bibinfo{volume}{73}}, \bibinfo{pages}{141--144}, \doiprefix\url{10.1080/00207219208925653} (\bibinfo{year}{2007}).

\bibitem{RN302}
\bibinfo{author}{Smith, D.~R.}
\newblock \bibinfo{journal}{\bibinfo{title}{Analytic expressions for the constitutive parameters of magnetoelectric metamaterials}}.
\newblock {\emph{\JournalTitle{Phys Rev E Stat Nonlin Soft Matter Phys}}} \textbf{\bibinfo{volume}{81}}, \bibinfo{pages}{036605}, \doiprefix\url{10.1103/PhysRevE.81.036605} (\bibinfo{year}{2010}).

\bibitem{RN303}
\bibinfo{author}{Smith, D.~R.}, \bibinfo{author}{Vier, D.~C.}, \bibinfo{author}{Koschny, T.} \& \bibinfo{author}{Soukoulis, C.~M.}
\newblock \bibinfo{journal}{\bibinfo{title}{Electromagnetic parameter retrieval from inhomogeneous metamaterials}}.
\newblock {\emph{\JournalTitle{Phys Rev E Stat Nonlin Soft Matter Phys}}} \textbf{\bibinfo{volume}{71}}, \bibinfo{pages}{036617}, \doiprefix\url{10.1103/PhysRevE.71.036617} (\bibinfo{year}{2005}).

\bibitem{RN851}
\bibinfo{author}{Shen, Q.} \emph{et~al.}
\newblock \bibinfo{journal}{\bibinfo{title}{Role played by port drains in a maxwell fish-eye lens}}.
\newblock {\emph{\JournalTitle{Journal of the Optical Society of America B}}} \textbf{\bibinfo{volume}{40}}, \bibinfo{pages}{1483}, \doiprefix\url{10.1364/josab.486187} (\bibinfo{year}{2023}).

\bibitem{RN827}
\bibinfo{author}{Frei, J.}, \bibinfo{author}{Xiao-Ding, C.} \& \bibinfo{author}{Muller, S.}
\newblock \bibinfo{journal}{\bibinfo{title}{Multiport $s$-parameter and $t$-parameter conversion with symmetry extension}}.
\newblock {\emph{\JournalTitle{IEEE Transactions on Microwave Theory and Techniques}}} \textbf{\bibinfo{volume}{56}}, \bibinfo{pages}{2493--2504}, \doiprefix\url{10.1109/tmtt.2008.2005873} (\bibinfo{year}{2008}).

\bibitem{pozar2011microwave}
\bibinfo{author}{Pozar, D.~M.}
\newblock \emph{\bibinfo{title}{Microwave engineering}} (\bibinfo{publisher}{John wiley \& sons}, \bibinfo{year}{2011}).

\bibitem{RN44}
\bibinfo{author}{Huang, H.} \& \bibinfo{author}{Shen, Z.}
\newblock \bibinfo{journal}{\bibinfo{title}{Low-rcs reflectarray with phase controllable absorptive frequency-selective reflector}}.
\newblock {\emph{\JournalTitle{IEEE Transactions on Antennas and Propagation}}} \textbf{\bibinfo{volume}{67}}, \bibinfo{pages}{190--198}, \doiprefix\url{10.1109/tap.2018.2876708} (\bibinfo{year}{2019}).

\bibitem{RN224}
\bibinfo{author}{Wu, Y.} \emph{et~al.}
\newblock \bibinfo{journal}{\bibinfo{title}{A broadband metamaterial absorber design using characteristic modes analysis}}.
\newblock {\emph{\JournalTitle{Journal of Applied Physics}}} \textbf{\bibinfo{volume}{129}}, \bibinfo{pages}{134902}, \doiprefix\url{10.1063/5.0043054} (\bibinfo{year}{2021}).

\bibitem{RN45}
\bibinfo{author}{Fan, C.} \emph{et~al.}
\newblock \bibinfo{journal}{\bibinfo{title}{Electromagnetic shielding and multi-beam radiation with high conductivity multilayer graphene film}}.
\newblock {\emph{\JournalTitle{Carbon}}} \textbf{\bibinfo{volume}{155}}, \bibinfo{pages}{506--513}, \doiprefix\url{10.1016/j.carbon.2019.09.019} (\bibinfo{year}{2019}).

\bibitem{RN46}
\bibinfo{author}{Wu, B.} \emph{et~al.}
\newblock \bibinfo{journal}{\bibinfo{title}{Broadband low-profile frequency selective rasorber using ultra-miniaturized metal-graphene structure}}.
\newblock {\emph{\JournalTitle{IEEE Antennas and Wireless Propagation Letters}}} \textbf{\bibinfo{volume}{21}}, \bibinfo{pages}{2422--2426}, \doiprefix\url{10.1109/lawp.2022.3195811} (\bibinfo{year}{2022}).

\end{thebibliography}

\end{document}